# Nanomolecular OLED Pixelization Enabling Electroluminescent Metasurfaces

Tommaso Marcato[1†], Jiwoo Oh[1†], Zhan-Hong Lin[1], Sunil B. Shivarudraiah[1], Sudhir Kumar[1], Shuangshuang Zeng[1,2]*, and Chih-Jen Shih[1]*

[1] Institute for Chemical and Bioengineering, ETH Zürich, 8093 Zürich, Switzerland

[2] College of Integrated Circuits, Huazhong University of Science and Technology, Wuhan, China

*All correspondence should be addressed to S.Z. (sszeng@hust.edu.cn) and C.J.S. (chih-jen.shih@chem.ethz.ch)

†These authors contributed equally.

**ABSTRACT**

Miniaturization of light-emitting diodes (LEDs) can enable high-resolution augmented and virtual reality displays[1,2] and on-chip light sources for ultra-broadband chiplet communication[3,4]. However, unlike silicon scaling in electronic integrated circuits[5], patterning of inorganic III-V semiconductors in LEDs considerably compromises device efficiencies at submicrometer scales[6–12]. Here, we present the scalable fabrication of nanoscale organic LEDs (nano-OLEDs), with the highest array density (>84,000 pixels per inch) and the smallest pixel size (~100 nm) ever reported to date. Direct nanomolecular patterning of organic semiconductors is realized by self-aligned evaporation through nanoapertures fabricated on a free-standing silicon nitride film adhering to the substrate. The average external quantum efficiencies (EQEs) extracted from a nano-OLED device of more than 4 megapixels reach up to 10%. At the subwavelength scale, individual pixels act as electroluminescent meta-atoms forming metasurfaces that directly convert electricity into modulated light. The diffractive coupling between nano-pixels enables control over the far-field emission properties, including directionality and polarization. The results presented here lay the foundation for bright surface light sources of dimension smaller than the Abbe diffraction limit[13], offering new technological platforms for super-resolution imaging, spectroscopy, sensing, and hybrid integrated photonics.

Miniaturization is a trend to manufacture ever-smaller devices and products. The exponential downscaling of silicon transistors, known as Moore's law[5], has led to the revolution of artificial intelligence in the past decade[14]. Driven by the technological development for high-speed data communication and processing, miniaturization of photonic components[15–17] such as light sources, waveguides and detectors, is at the forefront of research. Specifically for the light sources, emerging technologies such as augmented and virtual reality displays[1,2] and ultra-broadband on-chip communication[3,4,18,19] demand submicrometer scale LEDs operating in the visible and infrared wavelength regions[20–22], respectively. The inorganic III-V LED technology has been considered as a promising candidate. However, inorganic LEDs suffer from size-dependent EQE reduction, considerably compromising the electroluminescence (EL) efficiencies given the drastically increased non-radiative recombination defects induced during the semiconductor patterning process[6–12].

In this regard, a long-overlooked advantage of using organic emitting materials is the fact that the emission comes from Frenkel excitons highly localized within individual molecules in amorphous films[23], making OLED technology fundamentally in favor of miniaturization[24]. However, despite their prevalent application in large-area displays[25–27], the fabrication of small OLED pixels remains challenging, due to the incompatibility of organic materials with standard microfabrication processes based on photolithography[23]. For commercial OLED displays, patterning of organic semiconductors relies on the direct evaporation of organic molecules through the fine metal mask manufactured from Invar foil, which fails to achieve submicrometer resolution because of the large thickness of the mask.

Aimed at the broad applications of miniaturized LEDs, nanoscale patterning of organic semiconductors is indispensable. Here we present the scalable fabrication of nano-OLED pixels that achieves an array density of over 84,000 pixels per inch (ppi) and an individual pixel size of approximately 100 nm. Direct nanopatterning of solvent-sensitive organic emissive layer (EML) is realized through self-aligned nanostencil lithography. Nanostencil lithography[28–31] is a resistless technique involving shadow-mask evaporation through nanoapertures on a free-standing membrane, namely a nanostencil. The thinness of the nanostencil enables close adhesion to a target substrate, thereby allowing precise transfer of the nanoaperture patterns. The nanostencils can be reused multiple times after removing the materials deposited on them; a resist film can also be used to finely control the substrate-to-stencil gap, thus ensuring high yield and reproducibility[31].

In order to achieve the smallest OLED pixel size and spacing, nanostencil fabrication starts with the deposition of ultrathin (50 nm) silicon nitride ($SiN_x$) film grown on a silicon wafer employing low-pressure chemical vapor deposition (LPCVD). Nanoapertures in the $SiN_x$ film were patterned by electron-beam lithography, followed by selective reactive ion etching (RIE). The $SiN_x$ membrane was then released from the supporting substrate by a combination of dry and wet deep silicon etching. Details can be found in Methods and Supplementary Fig. S1. Figure 1a presents a 4-inch wafer containing 52 chips, where each chip has 36 freestanding windows comprised of nanostencil masks with different patterns.

We examined and optimized the SiN$_x$-based nanostencil lithography process by shadow-mask evaporating a phosphorescent guest:host EML system, 8 wt% Ir(ppy)3 doped in CBP onto silicon and glass substrates (for full compound names see Methods). Extended Data Figures 1b display representative scanning electron microscope (SEM) and atomic force microscope (AFM) images of the densest 2D and 1D periodic arrays of nanopatterned EML. The smallest pixel dimension is approximately 100 nm. As the nanoaperture dimension is comparable to the thickness of the SiN$_x$ membrane, we attain high degree of uniformity across the whole nanostencil chip area when aligning the substrate with the axis of the RADAK® source, thus ensuring perpendicular projection of the molecular beam.

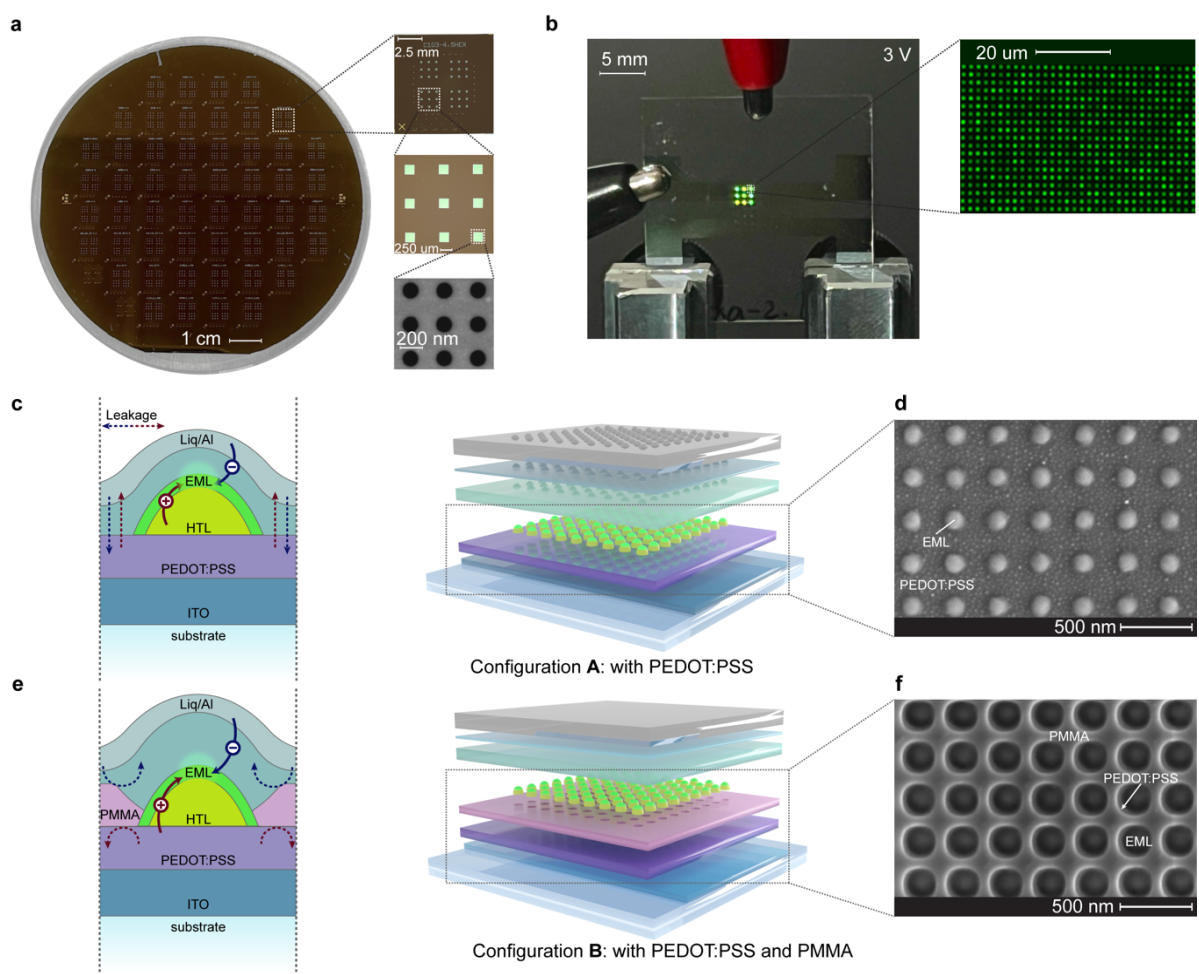

**Fig. 1: Nano-OLED technology. a,** Photographs of a 4-inch nanostencil wafer containing 52 chips. Each chip has 36 freestanding windows made by nanoaperture SiN$_x$ membrane of dimension of 200 × 200 μm$^2$. **b,** Photograph for an operating nano-OLED device at 3V, which contains > 4 megapixels of 2D hexagonal nanodisk arrays ($p$ = 350 nm; ~72,500 ppi). The inset shows a representative microscope image of an operating nano-OLED. Here to allow the individual pixels to be discerned, a 2D square nanodisk array with a larger spacing $p$ = 2

$\mu$m is chosen. **c,e,** Schematic diagrams for two device configurations, **A** and **B**, of our bottom-emitting nano-OLEDs fabricated on glass substrates, highlighting leakage current mitigation. The dashed lines represent electron (blue) and hole (red) leakage currents while solid lines are used for the recombination currents. **d,f,** Representative SEM images illustrating the device configurations, **A** and **B**. In particular, **f** shows the self-aligned overlay of the organic emitter patterns and the insulating PMMA barrier.

The optimized protocol for EML nanopatterning was incorporated in the fabrication of bottom-emitting nano-OLEDs on glass substrates based on the device architecture of ITO / PEDOT:PSS / TAPC / Ir(ppy)$_3$:CBP / B3PyMPM / Liq / Al (for full names of all materials see Methods). TAPC and B3PyMPM are the hole and electron transport layers (HTL and ETL), respectively. Our benchmark device involves sequential deposition of HTL and EML through the stencil chip onto ITO anode, followed by detaching the stencil chip prior to the evaporation of ETL, electron injection layer (EIL), and cathode (a schematic of the fabrication procedure is displayed in Extended Data Fig. 1a). Nevertheless, the device suffers from high leakage current due to direct contact between ETL and anode (Configuration **D** in Fig 2a). Several device structures were further examined, including Configurations **A** and **B**. (Fig. 1c-f). For the former, we mitigated the leakage current by using a low-conductivity PEDOT:PSS film as the hole injection layer (HIL) (cross-sectional transmission electron microscope (TEM) image and energy-dispersive X-ray spectroscopy (EDS) map see Extended Data Fig. 3c). In Configuration **B**, an insulating PMMA barrier layer was inserted between HIL and ETL, which was patterned by oxygen plasma etching through the stencil chip before HTL/EML deposition, yielding self-aligned nano-OLED pixels surrounded by insulating barriers (top-view SEM image in Fig. 1f and additional information in Extended Data Fig. 2).

The scalable fabrication of nano-OLEDs allows precise characterization of their EQEs using identical instrumentation to that of large-area planar LEDs. Figure 1b presents a representative photograph of our nano-OLED device for EQE measurement, which collects photons emitted from nano-OLED arrays of greater than 4 megapixels. The specific device in the photograph consists of hexagonal arrays of nanodisks (diameter of 100 nm and periodicity, $p$, of 350 nm). Extended Figure 2e shows the PL and EL micrographs for other nano-OLED designs of larger periodicity. The EL of nano-OLED shows less uniformity across the whole device area compared to PL, as the pixels sharing the same cathode have different values of electrical resistance.

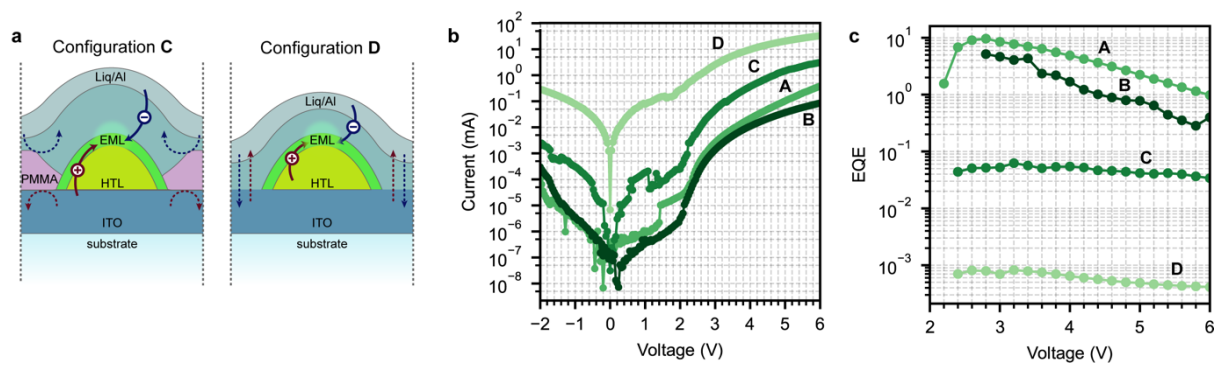

**Fig. 2: Nano-OLED EL characteristics. a,** Schematic drawings of the benchmark Nano-OLED configurations **C** and **D**. **b,c,** Representative nano-OLED device characteristics for current (**b**) and EQE (**c**) as a function of voltage. The highest EQE of approximately 10% is achieved in Configuration A.

We have evaluated device characteristics for four nano-OLED device configurations (Figs. 1c,e and 2a). Compared to Configuration **B**, Configurations **C** did not adopt the low-conductivity PEDOT:PSS layer. Figure 2b and 2c compare their current ($I$) and EQE responses with voltage ($V$). As expected, Configuration **B** suppressed the leakage current at the highest extent. At the device turn-on voltage (2.3 V), the current is reduced by 5 orders of magnitude as compared to Configuration **D**. However, we observed a trade-off between leakage current and large-area uniformity. In Configuration **B**, large-area overlay of insulating barrier and organic nanopatterns requires accurate alignment between EML molecular and oxygen plasma fluxes through the membrane nanoapertures. The narrow radial flux distribution of the RADAK® source lowers the yield over a large area, compromising the peak EQE (5.2%). In Configuration **A**, The peak EQE reaches 9.6% at 2.6 V; all EQE values are consistently above 1% throughout the voltage range considered here. We attribute the high efficiencies to the bottom-up nanopatterning of organic semiconductors which does not induce structural defects.

The nanoscale patterning of organic semiconductors enables unprecedentedly small dimensions of LED pixel size ($< \lambda/4$) and periodicity ($< \lambda/2$) at visible frequencies. The access to subwavelength scales shifts the LED operating regime to that of wave optics allowing the OLED pixels to act as meta-atoms for the first realization of electroluminescent organic metasurfaces. As demonstration we focus on the control of two crucial aspects of OLED emission for technological applications: EL directionality and polarization. For this purpose, we design two families of electroluminescent metasurfaces: 2D periodic arrays of nanodisks for directional emission and linear arrays of nanorods for the control of the degree of linear polarization of electroluminescence. For each nano-OLED array design, we examined the far-field focal plane images for pixelized EML deposited on bare glass substrate (PL) and bottom-emitting nano-OLED devices fabricated on glass (EL) using a high-numerical-aperture (high-NA) Fourier microscope[32,33].

The radiation generated by the nano-OLED arrays travels through space as spherical waves with a distinct angular distribution of intensity, polarization, and energy. Each angle up to the objective's NA, corresponding to an in-plane photon momentum $(k_x, k_y) = k_0(\sin\theta\cos\phi, \sin\theta\sin\phi)$, where $k_0 = 2\pi/\lambda$ is the free-space momentum of photon with wavelength $\lambda$, is collected and mapped onto the back focal plane (BFP) of microscope objective, generating the BFP image effectively equivalent to the spatial Fourier transform of the front focal plane image (see Methods).

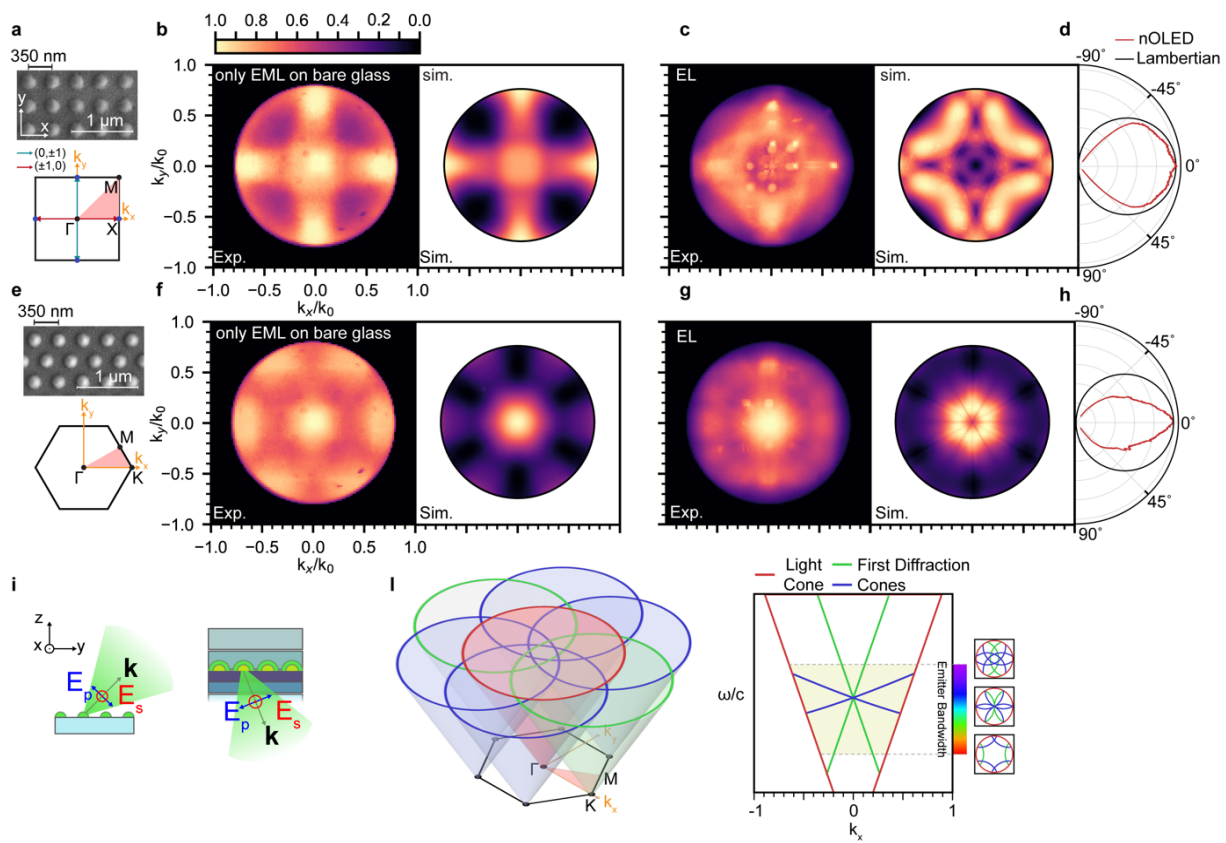

**Fig. 3: Controlling emission directionality of electroluminescent metasurfaces made by 2D arrays of nanodisk OLED pixels. a-e,** SEM images and the first Brillouin zones for 2D square and hexagonal arrays of EML nanodisks deposited on glass (**a** and **e**). **b,f,** The corresponding experimental (exp.) and simulated (sim.) back focal plane images (**b** and **f**)

directly reflect the rotational symmetries of the point group of the reciprocal lattice. **c,g,** EL back focal plane images for the corresponding electroluminescent metasurface, revealing that the square array directs light to high angles ($0.3 < |k_\parallel| < 0.8$) and the hexagonal array concentrates light at low angles ($|k_\parallel| < 0.25$). **d,h,** Polar plots of the electroluminescent metasurfaces angle dependent EL obtained as vertical cuts in the back focal plane images corrected by the cos$\theta$ apodization factor. Black lines illustrate the comparison with the Lambertian profile expected from traditional OLEDs with uniform EML. **i,** Schematic diagrams showing experimental frames of reference with respect to the array axis and the directions of polarizer slit for optical characterization. **l,** Schematic illustration for the reciprocal lattice of the electroluminescent metasurfaces and the resulting radiation pattern. The response of the array in *k*-space is equal to the convolution of the free photon dispersion (the light cone), and the reciprocal lattice (the intersection of light cones centered at the reciprocal lattice points). Each frequency component of the emitter spectrum contributes with an horizontal isofrequency slice of the array response, resulting in a broadening of the sharp circular modes. To the right the energy-momentum dispersion is represented up to first order diffraction. Depending on the angle between the grating vector **G** and the k$_x$ axis, the vertical ω-k$_x$ plane cuts the cones in the lattice dispersion resulting in either linear or hyperbolic photonic bands.

For the first family of electroluminescent metasurfaces we design the lattice periodicity $p$ to match the Ir(ppy)$_3$ emission maximum to the second Bragg condition of the lowest waveguided mode of the OLED stack, where the local density of states at the Γ point, $|k_\parallel| = 0$ (normal incidence) is maximized. To decouple the BFP images from the contributions of OLED dielectric stack, we first analyzed PL from the metasurfaces comprising nanodisk arrays of EML deposited on a bare glass substrate (reference frame see Fig. 3i left). The nanodisk arrays can be modeled as square and hexagonal photonic crystal lattices having $x - y$ periodic modulation of refractive index (SEM images and Brillouin zones see Figs. 3a and 3e), which determines the electromagnetic Bloch modes in the associated photonic band structures. The nanodisks of emitting material radiate incoherently in a broad range of wavevectors coupling to the Bloch modes under phase matching conditions, in which a wavevector is conserved modulo the reciprocal lattice vector $G = 2\pi/p$, or the "grating" momentum, following $|k_\parallel| = |k_\parallel^{in} + G|$, where $k_\parallel$ and $k_\parallel^{in}$ are the wavevectors of output and input modes, respectively. Accordingly, the reciprocal lattice of the pixel arrays modulates the radiation pattern that couples to free-space photons following $|k_\parallel| < k_0$ (see Fig. 3l)[34]. One can see the feature in their PL BFP images (Figs. 3b and 3e), whose rotational symmetries directly reflect the point groups of the corresponding reciprocal lattices.

The EL BFP images for the full nano-OLEDs devices (Figs 3c and 3g) involve complications given the existence of plasmonic and waveguide modes (reference frame see Fig. 3i right). These modes, which are otherwise not accessible, can be diffracted into free space photons because the EML periodic nanopatterns are transferred to the ETL and Al layers (Extended Data Fig. 2). The angle-dependent PL spectra in Extended Data Fig. 4 allow us to distinguish three distinct modes through their characteristic dispersions. The spectra show clear Fano resonances where the broad emission spectrum of the EML couples to sharp waveguided modes. The device supports two transverse magnetic (TM), or p-polarized, modes: a broader resonance at lower frequency consistent with a plasmonic mode of the Al cathode and a narrow $TM_0$ ITO waveguided mode, which according to the electric field profiles (Supplementary Fig. S5) are hybridized into waveguide-plasmon polaritons[35,36]. The s-polarized mode corresponds to the transverse electric ($TE_0$) ITO waveguided mode.

Remarkably, despite the broadband nature of the emissive spectrum, the simulated and experimental BFP images for the device stack (Fig. 3f) reveal that the majority of radiation is concentrated within $|k_\parallel| < 0.25 k_0$, or $\pm 15°$ in air. On the other hand, the electroluminescent metasurface of square arrays (Figs. 3a-c) directs light to higher angles, $0.3 < |k_\parallel| < 0.8$, because the modes coupled to air are moved towards the X point. The proof-of-concept demonstration showcases that engineering the lattice symmetry of nanodisk arrays alone can already redistribute light to different angles, in stark contrast to typical wide-angle Lambertian radiation generated from planar OLEDs (see Figs. 3d and 3h)[23].

The second family of electroluminescent metasurfaces that we investigated are based on a nanorod meta-atom having emission preferentially polarized along its long axis at all angles (see Supplementary Discussion 1). For its complete characterization, we implemented angle-dependent polarimetry[37,38] by adding a linear polarizer and quarter-wave plate before the Fourier lens, so that the four Stokes parameters, namely, $S0$, $S1$, $S2$, and $S3$, at each $(k_x, k_y)$ can be extracted to inform the polarization states of light in the momentum space (details see Methods). The *s*- and *p*-polarization correspond to the directions in parallel and orthogonal to the nanorod axial axis, respectively. In particular, the parameter $S0$ measures the total field intensity, and the parameter $S1$ reflects the degree of linear polarization with respect to the basis parallel ($k_x$) and perpendicular ($k_y$) following $S1(k_x, k_y) = I^s(k_x, k_y) - I^p(k_x, k_y)$, where $I^s$ and $I^p$ are the emission intensity at a given momentum extracted from the *s*- and *p*-polarized BFP images, respectively. The ratio $S1/S0$ quantifies the degree of linear polarization, ranging from +1 to -1. A more positive (negative) $S1/S0$ value denotes the light is more linearly polarized along the parallel (perpendicular) basis; unpolarized light yields $S1/S0 = 0$ [37,38].

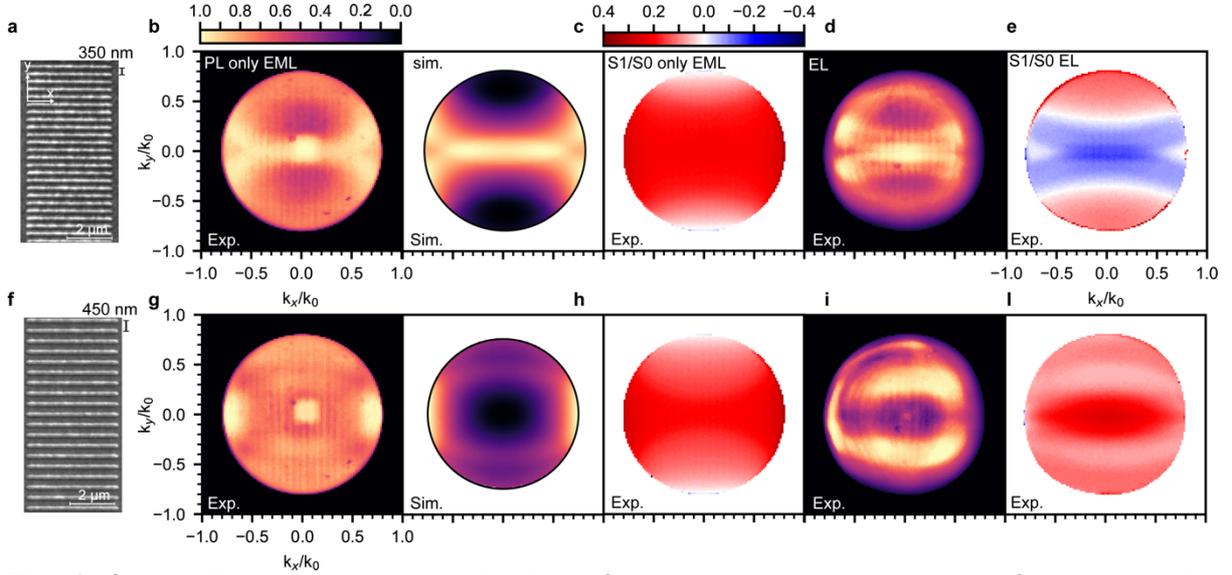

**Fig. 4: Controlling emission polarization of electroluminescent metasurfaces made by 1D arrays of nanorod OLED pixels. a-l,** SEM images for 1D arrays of EML nanorods deposited on glass with $p$ of 350 (**a**) and 450 nm (**f**). The corresponding experimental (exp.) and simulated (sim.) radiation patterns (**b** and **g**) reveal emission directionality along $k_y = 0$ axis, and the measured $S_1/S_0$ maps show *s*-polarized PL at all momenta (**c** and **h**). The EL radiation patterns (**d** and **i**) and $S_1/S_0$ maps for the nano-OLED devices (**e** and **l**) reveal that the EL polarization is strongly influenced by the leaky waveguide and plasmonic modes, such that the $p = 350$ nm nanorod arrays yield *p*-polarized light ($S_1/S_0 < 0$) at low $k_y$, whereas the $p = 450$ nm nanorod arrays remain to yield *s*-polarized light ($S_1/S_0 > 0$) at all angles. A change of nanorod periodicity reverses the electroluminescent polarization at $k = 0$.

Analogous to the 2D nanodisk arrays discussed earlier, even though the emission generated from nanopatterned EML is completely incoherent, the BFP images for the 1D arrays reveal that light is directed along $k_y = 0$ axis, orthogonal to the periodic direction *via* coherent scattering (Figs. 4b,d and 4g,i). The corresponding $S_1/S_0$ PL maps (Figs. 4c and 4h) exhibit preferentially parallel polarization at all angles. A lack of dependence on the array periodicity in the $S_1/S_0$ maps confirms that the observed linear polarization is dominated by the response of a single nanorod.

In the nano-OLED devices, the polarization of leaky waveguide ($TM_0$ and $TE_0$) and plasmonic TM modes comes into play (Figs. 4e and 4l). As a result, the EL polarization response of 1D nano-OLEDs strongly depends on the array periodicity. In particular, the electroluminescent metasurface made by nanorod arrays of $p$ = 350 nm exhibits the opposite (perpendicular, or $p$-pol., $S_1/S_0 < 0$) polarization for $|k_y| < 0.3$, due to the polarization of the localized plasmon resonance. A high degree of perpendicular polarization ($S_1/S_0$ = -0.25) is attained at zero angle, $k = 0$, where the EL is the strongest. On the other hand, the electroluminescent metasurface made by nanorod arrays of $p$ = 450 nm yields a strong preference of parallel polarization ($s$-pol., $S_1/S_0 > 0$) at all momenta. A highly parallel-polarized light with $S_1/S_0$ of 0.4 is achieved at $k = 0$. Remarkably, the variation in the periodicity of 1D nano-OLEDs can reverse the polarization of far-field electroluminescence.

In summary, we have demonstrated scalable fabrication of nano-OLEDs with pixel size and periodicity smaller than the diffraction limit of visible light without compromising device efficiency. Direct nanopatterning of emissive organic semiconductors gives rise to the electroluminescent metasurfaces that convert electricity to modulated light with controlled directionality and polarization. We anticipate the capability for tailoring the photonic landscape of organic semiconductors will open a new degree of freedom for LED design, offering extended benefit in emerging OLED technologies such as polariton OLEDs[39,40].

## METHODS

### Nanostencil fabrication

A double-side polished 300 μm-thick 4-inch Si wafer was used as a substrate (Fig S1a). A uniform layer of low-stress $SiN_x$ (thickness 50nm) was deposited (Fig S1b) using LPCVD (PEO-604, ATV Technologies GmbH).

EBL was used to define nanopatterns on the $SiN_x$ membrane (Fig S1c). An e-beam resist (AR-P 6200.09, Allresist GmbH) was spin coated, followed by e-beam exposure. Development was done using amyl acetate (AR 600-546, Allresist GmbH) as developer and DI water as stopper. After development, the substrate is soft baked for 2 minutes at 130 °C. Using e-beam resist as mask, $SiN_x$ is dry etched (Fig S1d) in a RIE chamber (PlasmaPro 80 RIE, Oxford Instruments). Anisol was used as a solvent for e-beam resist strip.

Photolithography was used to define substrate back-side opening area (Fig S1e). After coating a photoresist (AZ® 10XT 520CP, Microchemicals GmbH), PR exposure was done by back-side alignment with a premade photomask using a mask aligner (EVG®620 NT, EV Group). PR development was done using AZ® 400K 1:4 (Microchemicals GmbH) as a developer and DI water as stopper. Development was followed by a hard bake of 10 minutes at 115 °C. Before the back-side silicon etching, $SiN_x$ membrane on the back side was etched using RIE (Fig S1f).

Bosch process is used for deep silicon etching in a Si DRIE chamber (Omega® Rapier, SPTS). PR was used as a mask. About 260 out of 300 μm of the Si wafer was etched using DRIE. Remaining silicon wafer was etched using KOH wet etching. The wet etching was done in a water bath at 80 °C (Fig S1g).

### Device fabrication

Following materials were used for device fabrication : indium tin oxide (ITO), Poly(3,4-ethylenedioxythiophene)-poly(styrenesulfonate) (PEDOT:PSS), Poly(methyl methacrylate) (PMMA), Di-[4-(N,N-di-p-tolyl-amino)-phenyl]cyclohexane (TAPC), Tris(2-phenylpyridine)iridium(III) (Ir(ppy)3), 4,4'-Bis(N-carbazolyl)-1,1'-biphenyl (CBP), 4,6-Bis(3,5-di(pyridin-3-yl)phenyl)-2-methylpyrimidine, 4,6-Bis(3,5-di-3-pyridinylphenyl)-2-methylpyrimidine (B3PymPm), 8-Hydroxyquinolinolato-lithium (Liq), and aluminium (Al).

Following procedure describes fabrication of device in configuration B shown in fig 1. For other configurations, one or both spin-coating layers are omitted, while the same thermal evaporation steps apply.

An ITO-glass substrate (ITO sheet resistance = 10-15 Ω/sq) was cleaned in acetone and isopropanol, followed by oxygen plasma treatment at 200W for 10 minutes (Plasmalab80Plus, Oxford Instruments). Microfiltered PEDOT:PSS solution was spin-coated as a hole injection layer (Clevios™ P VP CH 8000, Heraeus GmbH), baked at 120°C for 15 minutes. A stencil chip is attached on the substrate with alignment to the pre-patterned ITO electrodes. For devices with insulation layer, a thin PMMA layer was spin-coated and baked at 180 °C for 5 minutes. Oxygen plasma etching was used to etch the PMMA layer through the nanostencil and expose the underlying PEDOT:PSS layer. Following organic layers were thermally evaporated through the nanostencil: TAPC (hole transport, 40 nm) / Ir(ppy)$_3$:CBP (emission, 20 nm). Then, the nanostencil is detached from the substrate. The following layers were thermally evaporated through an open shadow mask : B3PyMPM (electron transport, 50nm) / Liq (electron injection, 2 nm) / Al (cathode, 70 nm).

**Nano-OLED morphological and topographical characterization**

SEM (Scios™ 2 DualBeam™, Thermo Fisher Scientific) images were taken to characterize morphology of nanostructures, optimize etching parameters and large area uniformity. AFM (Park NX10, Park Systems) was used to acquire high resolution data of nanostructures and topographical characteristics. The measurements were done in non-contact mode using a force modulation probe (PPP-FMR, Nanosensors™). TEM (Talos F200X, Thermo Fisher Scientific) measurements were performed in dark and bright field modes to investigate the structural information of the cross-section of nano-OLED devices. Focused ion beam (FIB, Hellios NanoLab 600i, Thermo Fisher Scientific) technique was used to prepare lamella sample from the nano-OLED device for TEM measurement. Protective carbon layer was deposited on top of the nanopattern. Then, Ga liquid ion source was used to mill the area around the protection layer. EDS measurements were performed using the same instrument as TEM but using the Super-X EDS detector.

**Device characterization**

The I-V, luminance and EQE characteristics are measured using two source-measurements units (SMU, Keithley 2450, Tektronix) and a large area (10 mm x 10 mm) calibrated Si photodiode (FDS1010-CAL, Thorlabs) placed in direct contact with the bottom surface of the device substrate to ensure underfilling of the detector and the collection of all the photons in the forward direction[41] (see Supplementary Fig. S3a). The electroluminescence (EL) spectra are measured with a spectroradiometer (PR 655 SpectraScan, Photo Research) or an imaging spectrograph (Kymera 328i, Andor).

**Optical characterization**

A schematic of the home-built optical characterization setup is displayed in Supplementary Figure S3b,c. EML nanostructure arrays and Nano-OLEDs are measured with an inverted microscope (Eclipse Ti2-U) equipped with a 50X dry objective (TU Plan Fluor, NA = 0.8, WD = 1.00 mm). For PL measurements the samples are excited with widefield illumination with a 400 nm LED source (pE-300, CoolLED). The emission is collected by the same objective and sent through a dichroic beamsplitter (LF-405/LP-B filter cube, Semrock) and to the exit port of the microscope where an iris in the image plane is used to select the region of interest. An achromatic doublet (AC254-200-A, f = 200 mm same as the microscope's tube lens, Thorlabs) placed with the image plane in its front focal plane produces the desired k-space image onto its back focal plane where an imaging spectrograph (Kymera 328i, Andor) with its slit wide open relays the image to a sCMOS camera (Zyla 4.2 PLUS, Andor). For imaging a mirror is used instead of a zero-order grating to avoid aberrations. The same setup can real space images onto the camera by placing a 100 mm lens halfway between the image plane and the Fourier lens. For EL measurements the devices are contacted from the top with tungsten probes (Signatone) connected to an SMU (Keithley 2450, Tektronix). Angle-resolved PL, EL and reflectivity spectra are obtained by closing the spectrometer slit onto the k-space image and the image is dispersed by a grating (150 lines mm$^{-1}$ blazed at 500 nm) in the direction orthogonal to the slit. The sample is rotated to ensure that the $k_y$-axis of the patterns is parallel to the slit. A linear polarizer is inserted before the Fourier lens with its axis parallel or orthogonal to the spectrometer's entrance slit to obtain the angle dependent second Stokes parameter S1 and polarized spectra.

**Optical simulations**

Numerical simulations were performed using Lumerical (Ansys Inc.). The farfield radiation patterns from periodic structures were calculated using the rigorous coupled wave approximation (RCWA). The device unit cell was illuminated by a plane wave of wavelength λ, incident angle $(\theta, \phi)$ and s or p polarization. The polar angle $\theta$ was varied from 0° up to 53°, corresponding to the objective's NA, while the azimuthal angle from 0° to 360°. A small imaginary part (0.001i) was added to the refractive index of the EML to ensure the reciprocity principle could be used to estimate the emission into the farfield[42,43]. All the other dielectric materials were considered lossless and dispersion was accounted for by using wavelength-dependent refractive index. The procedure yields the transmitted and reflected field intensities which can be combined to calculate the resulting absorption by energy conservation. The final farfield radiation pattern is obtained by integrating the absorption weighted by the PL or EL spectrum and by averaging the two orthogonal polarizations. The field profiles in Supplementary Fig. S5 are obtained from the same simulation employing the finite-difference time-domain (FDTD) method with plane wave illumination at normal incidence.

## ACKNOWLEDGMENTS

The authors are grateful for financial support from following projects: ETH Grant (ETH-33 18-2), Swiss National Science Foundation (Project 200021-178944), and European Research Council (Starting Grant # 849229 CQWLED). The authors also appreciate strong technical support from (i) microfabrication facilities at the Center for Micro- and Nanoscience at ETH Hönggerberg (FIRST) and the Binnig and Rohrer Nanotechnology Center at IBM Research Zurich (BRNC), and (ii) electron microscopy facilities at the Scientific Center for Optical and Electron Microscopy (ScopeM).

## AUTHOR CONTRIBUTIONS

T.M., J.O., S.Z., and C.J.S. conceived the idea and designed the experiments. T.M. built the Fourier microscope and performed optical simulations under advice from Z.L. T.M. carried out optical and EL characterization. S.Z. established protocols for the fabrication of nanostencils. T.M. established protocols for the fabrication of nano-OLED devices. J.O. designed and fabricated nanostencils and nano-OLED devices. J.O. carried out SEM and AFM characterization. S.B.S prepared lamella by FIB milling and performed cross-sectional TEM imaging and EDS. S.K. established protocols and device architecture for the fabrication of reference OLED devices. T.M., J.O., and C.J.S. co-wrote the paper. All authors contributed to this work, commented on the paper, and agreed to the contents of the paper and supplementary materials.

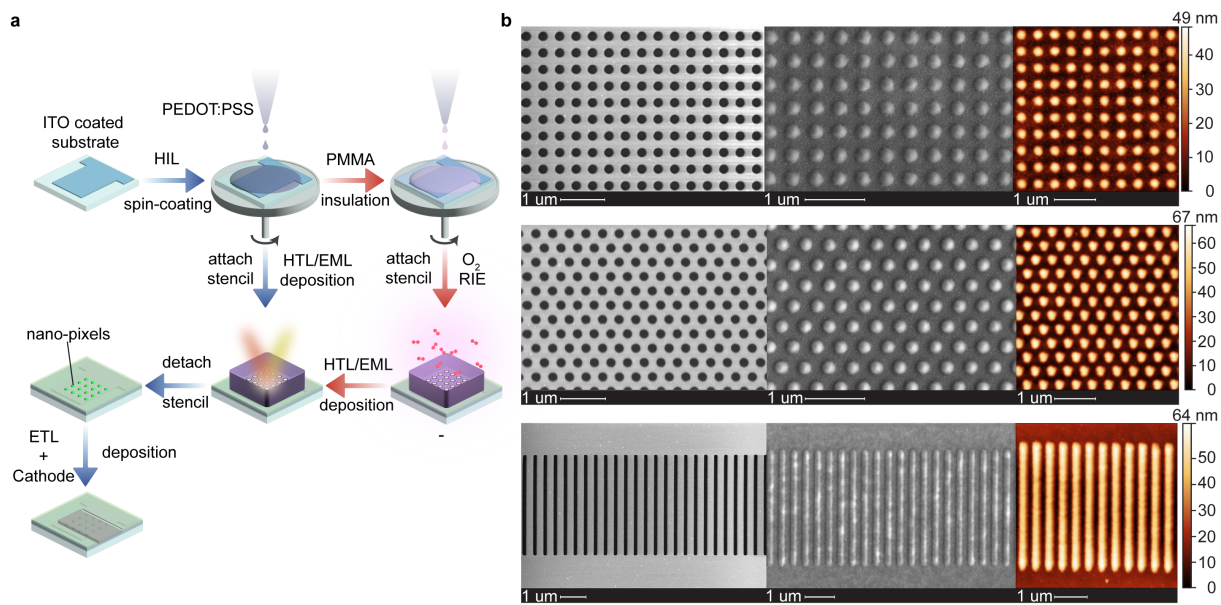

**Extended Data Fig. 1. | Nano-OLED fabrication a,** Schematic diagram illustrating the fabrication process for the bottom-emitting Nano-OLEDs. The additional pathway following the red arrows leads from Configuration **A** to **B. b,** Representative SEM and AFM images for some of the densest Ir(ppy)$_3$:CBP nanopixel arrays and the corresponding nanostencils used for their deposition. From top to bottom: 2D square and hexagonal arrays of 100 nm nanodisks 1D linear array of 0.1 µm x 4 µm nanorods with a periodicity of 300 nm.

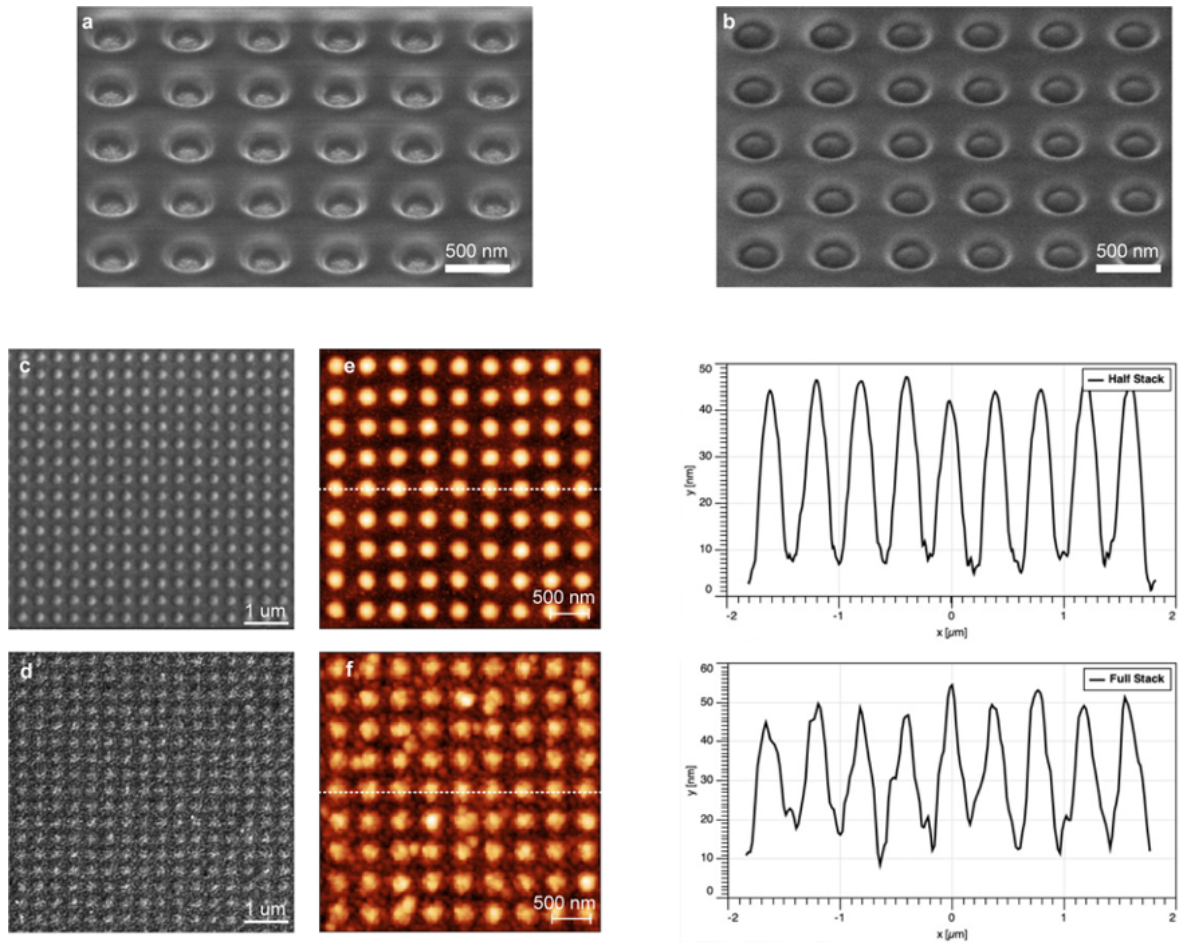

**Extended Data Fig. 2 | Nano-OLED stack morphology a,** SEM image of the insulating PMMA barrier with holes etched with O2 plasma through nanostencil. **b,** SEM image of a Nano-OLED half-stack in Configuration **B** where the overlay between HTL/EML nanodisks and PMMA holes is displayed. **c-f,** Representative SEM and AFM images of Nano-OLED half-stack (**c** and **e**) full-stack (**d** and **f**) illustrating the transfer of the EML pattern to the top Al cathode layer. Height profiles corresponding to the dashed linecuts are shown on the right. All images are taken from square 2D arrays of nanodisks with 200 nm diameter and 500 nm (**a** and **b)** or 400 nm **(c-f)** spacing.

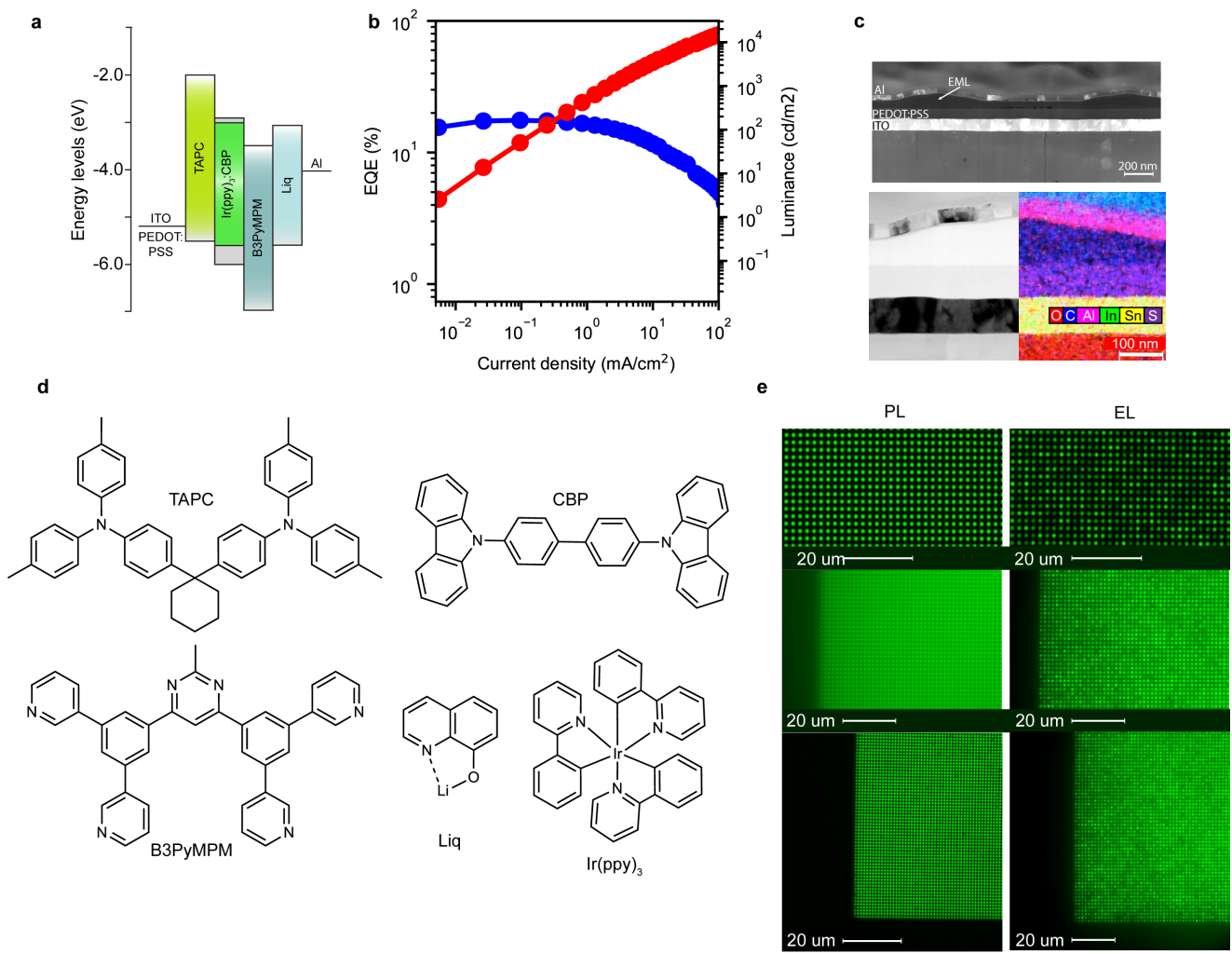

**Extended Data Fig. 3 | OLED reference device and architecture a,** Schematic drawing of the OLED energy level alignment **b,** EQE (blue) and luminance (red) characteristics as a function of current density of the reference OLED with uniform EML. The peak EQE is 17.3 ± 0.3%. **c,** Cross-sectional TEM images and EDS map of a representative Nano-OLED device consisting of a 1D linear array of 300 nm wide nanorods with 1 μm periodicity. **d,** Chemical structures of the materials used in our optimized OLED architecture. **e,** PL (left) and EL (right) OM images for 2D square nanodisk arrays with periodicity, from top to bottom, of $p =$ 2 μm, 1.5 μm and 1 μm operating at 5 V.

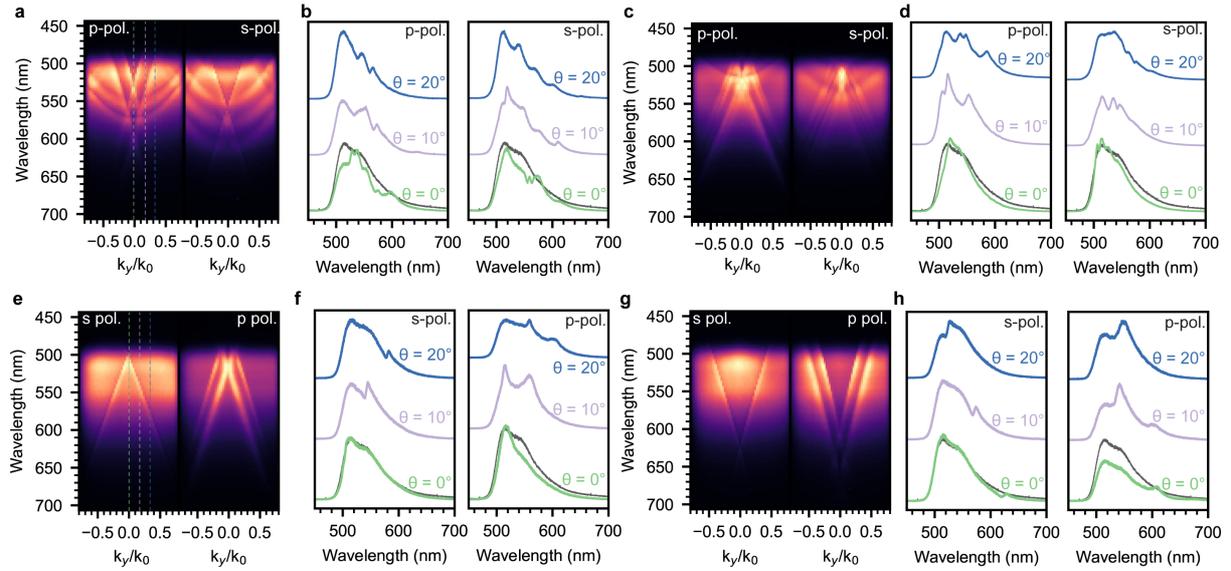

**Extended Data Fig. 4 | Angle-dependent spectra of electroluminescent metasurfaces made by arrays of nano-OLED pixels a-h** Angle-resolved PL spectra for the devices in Fig. 3 and 4 having 2D square (**a**), hexagonal (**c**) nanodisks and 1D nanorods (**e** and **g**) pixel arrays along the $\Gamma - X$, $\Gamma - K$ and $k_y$ directions in the Brillouin zone, respectively, with the polarizer parallel (*p*-pol.) and perpendicular (*s*-pol) to the spectrometer slit. Spectra corresponding to line cuts at 0º, 10º, and 20º (**b, d, f** and **h**) were extracted to resolve the TM (*p*-polarized; *p*-pol) and TE (*s*-polarized; *s*-pol) modes. The gray curves correspond to PL spectra for the uniform EML film.

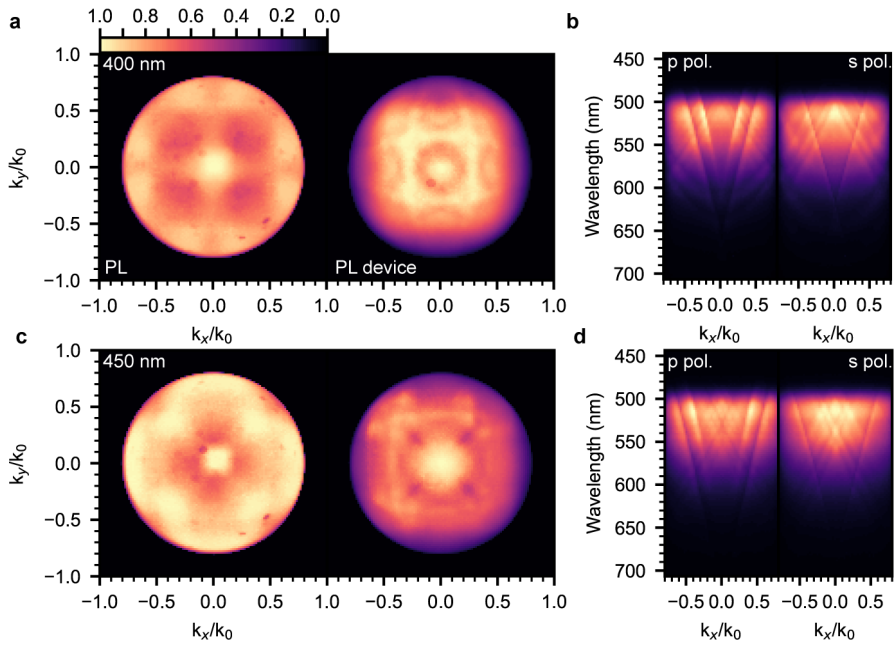

**Extended Data Fig. 5 | PL characteristics for electroluminescent metasurfaces based on square arrays of nanodisk OLED pixels. a,c** Experimental PL back focal plane images for square nanodisk arrays on glass and their corresponding electroluminescent metasurfaces with $p = 400$ nm (**a**) and 450 nm (**c**). **b,d** Corresponding angle-dependent PL spectra along $\Gamma - \mathrm{X}$ direction with the polarizer parallel (*p*-pol.) and perpendicular (*s*-pol) to the spectrometer slit.

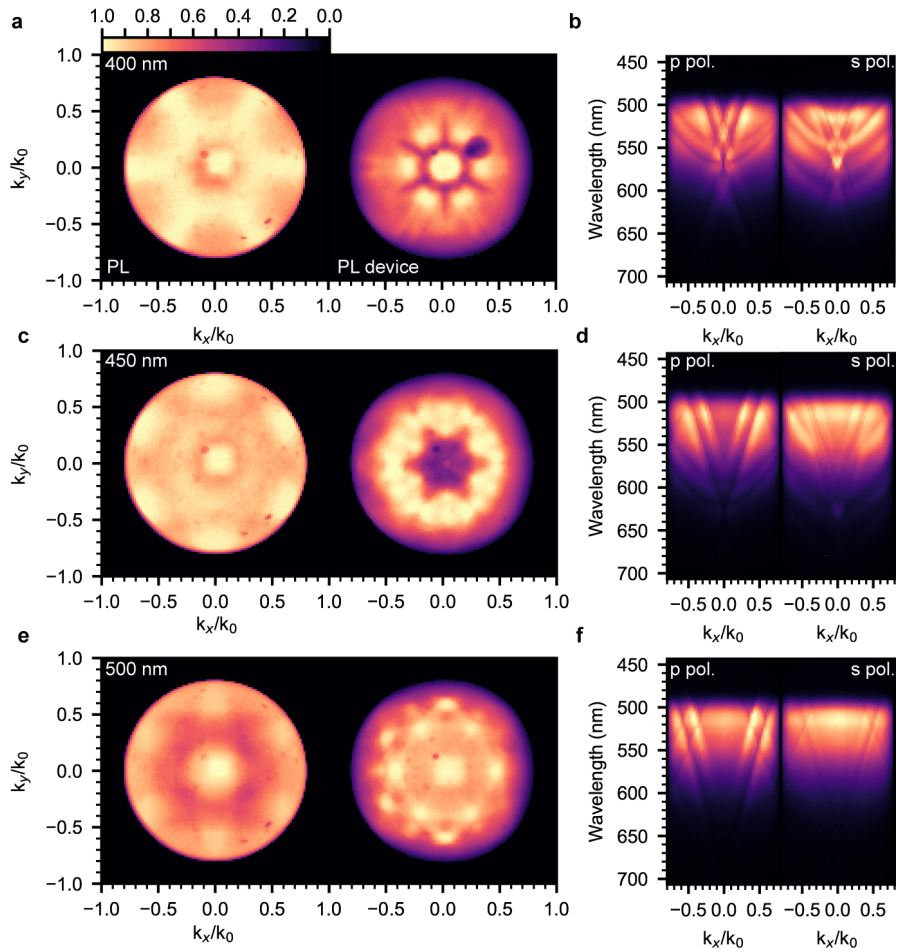

**Extended Data Fig. 5 | PL characteristics of electroluminescent metasurfaces based on hexagonal arrays of nanodisk OLED pixels a,c,e** Experimental PL back focal plane images of hexagonal nanodisk arrays of 200 nm diameter on glass and their corresponding electroluminescent metasurfaces with $p = 400$ nm (**a**), 450 nm (**c**) and 500 nm (**e**) spacing. **b,d,f** Corresponding angle-dependent PL spectra along $\Gamma - X$ direction with the polarizer parallel (*p*-pol.) and perpendicular (*s*-pol) to the spectrometer slit.

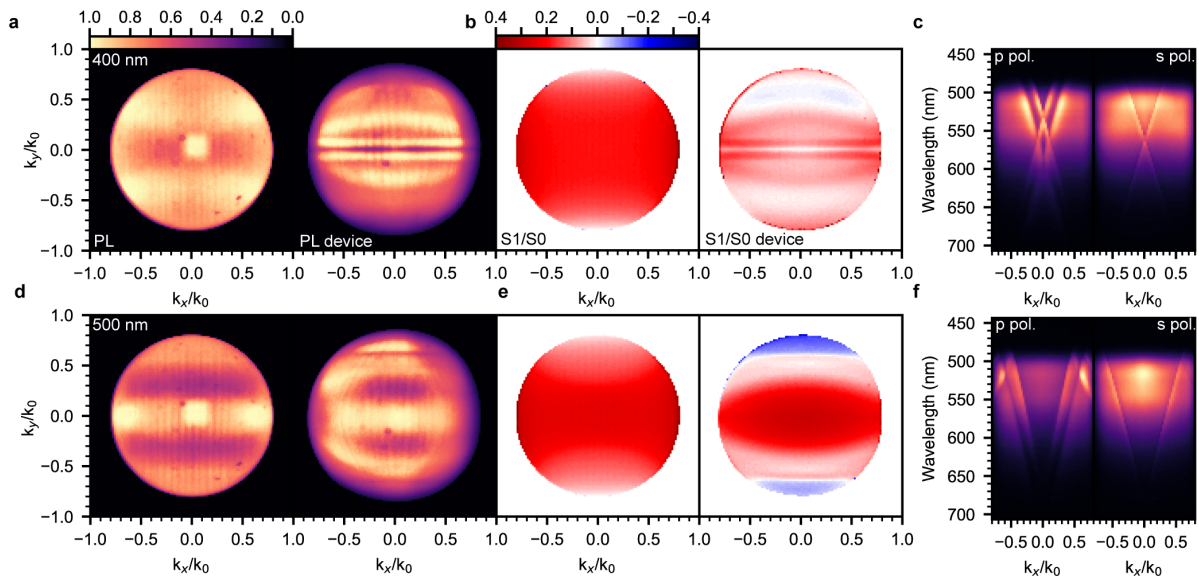

**Extended Data Fig. 6 | PL and polarization characteristics of electroluminescent metasurfaces based on 1D arrays of nanorod OLED pixels a,d** Experimental PL back focal plane images of 1D arrays of 200 nm width on glass and their corresponding electroluminescent metasurfaces with 400 nm **(a)** and 500 nm **(d)** spacing. **b,e** Corresponding S1/S0 linear polarization maps. **c,f,** Corresponding angle-dependent PL spectra along $\Gamma - X$ direction with the polarizer parallel (*p*-pol.) and perpendicular (*s*-pol) to the spectrometer slit.